\documentclass[prb,aps,epsf,twocolumn,showpacs,superscriptaddress]{revtex4}
\usepackage{graphicx}
\usepackage{latexsym}
\usepackage{amsmath}
\begin{document}
%\preprint{cond-mat/0305xxx}\tolerance = 10000
%\documentstyle[prb,aps,epsf,floats]{revtex}
%\documentstyle[pra,aps,epsf,floats]{revtex}
\newcommand{\beq}{\begin{equation}}
\newcommand{\beqr}{\begin{eqnarray}}
\newcommand{\eeqr}{\end{eqnarray}}
\newcommand{\eeq}{\end{equation}}
\newcommand{\s}{{\sigma}}
\newcommand{\e}{{\varepsilon}}
\newcommand{\om}{{\omega}}
\newcommand{\Om}{{\Omega}}
\newcommand{\D}{{\Delta}}
\newcommand{\de}{{\delta}}
\newcommand{\al}{{\alpha}}
\newcommand{\ga}{{\gamma}}
\newcommand{\La}{{\Lambda}}
\newcommand{\be}{{\beta}}
\newcommand{\psib}{{\bar{\psi}}}
\newcommand{\rb}{{\bar{\rho}}}
\newcommand{\phib}{{\bar{\phi}}}
\newcommand{\dt}{{\Delta}}
\newcommand{\dva}{{\frac{\vp\times\va}{2\pi}-\rb}}
\newcommand{\dvab}{{\frac{\vp\times(\va-\vb)}{4p\pi}}}
\newcommand{\w}{{\omega}}
\newcommand{\zh}{{\hat{z}}}
\newcommand{\qh}{{\hat{q}}}
\newcommand{\vA}{{\vec{A}}}
\newcommand{\va}{{\vec{a}}}
\newcommand{\vrr}{{\vec{r}}}
\newcommand{\vj}{{\vec{j}}}
\newcommand{\vE}{{\vec{E}}}
\newcommand{\vB}{{\vec{B}}}
\def\bA{{\mathbf A}}
\def\bm{{\mathbf m}}
\def\bsig{{\mathbf \sigma}}
\def\bB{{\mathbf B}}
\def\bp{{\mathbf p}}
\def\bI{{\mathbf I}}
\def\bn{{\mathbf n}}
\def\bM{{\mathbf M}}
\def\bq{{\mathbf q}}
\def\bp{{\mathbf p}}
\def\br{{\mathbf r}}
\def\bs{{\mathbf s}}
\def\bS{{\mathbf S}}
\def\bQ{{\mathbf Q}}
\def\bs{{\mathbf s}}
\def\bB{{\mathbf B}}
\def\bl{{\mathbf l}}
\def\bPi{{\mathbf \Pi}}
\def\bJ{{\mathbf J}}
\def\bR{{\mathbf R}}
\def\bz{{\mathbf z}}
\def\ba{{\mathbf a}}
\def\bk{{\mathbf k}}
\def\bK{{\mathbf K}}
\def\bP{{\mathbf P}}
\def\bg{{\mathbf g}}
\def\bX{{\mathbf X}}
\newcommand{\Psib}{\mbox{\boldmath $\Psi $}}
\newcommand{\thetab}{\mbox{\boldmath $\theta $}}
\newcommand{\sigmab}{\mbox{\boldmath $\sigma $}}
\newcommand{\gammab}{\mbox{\boldmath $\gamma $}}
\newcommand{\vx}{{\vec{x}}}
\newcommand{\vq}{{\vec{q}}}
\newcommand{\vQ}{{\vec{Q}}}
\newcommand{\vd}{{\vec{d}}}
\newcommand{\vb}{{\vec{b}}}
\newcommand{\vp}{{\vec{\partial}}}
\newcommand{\p}{{\partial}}
\newcommand{\gr}{{\nabla}}
\newcommand{\ra}{{\rightarrow}}
\def\dd{d^{\dagger}}
\def\half{{1\over2}}
\def\third{{1\over3}}
\def\twof{{2\over5}}
\def\threes{{3\over7}}
\def\rhob{{\bar \rho}}
\def\ua{\uparrow}
\def\da{\downarrow}
\def\eqa{\begin{eqnarray}}
\def\eea{\end{eqnarray}}
\parindent=4mm
\addtolength{\textheight}{0.9truecm}
%\begin{document}
%\draft \flushbottom \twocolumn[
%\hsize\textwidth\columnwidth\hsize\csname
%@twocolumnfalse\endcsname
\title{Equality of bulk wave functions and edge correlations in  topological superconductors:\\
 A spacetime derivation.}

\author{R. Shankar}
\affiliation{Department of Physics, Yale University, New Haven CT 06520}
\author{Ashvin Vishwanath }
\affiliation{ Department of Physics, University of California, Berkeley CA 94720}
\date{\today}

%\tightenlines
%\widetext
%\advance\leftskip by 57pt
%\advance\rightskip by 57pt
\begin{abstract}

For certain systems, the N-particle ground-state wavefunctions of the bulk
happen to be exactly equal to the N-point space-time correlation functions
at the edge, in the infrared limit. We show why this had to be so for a
class of topological superconductors, beginning with the p+ip state in
D=2+1. Varying the chemical potential as a function of Euclidean {\em time} between
weak and strong pairing states is shown to extract the wavefunction.
Then a Euclidean rotation that exchanges time and space  and approximate Lorentz invariance
lead to the edge connection. We illustrate straightforward extension to other dimensions (eg. ${}^3$He- B phase in D=3+1) and to correlated states like fractionalized topological superconductors.

 %  We consider here two examples, the $p+ip$ superconductor  in $d=2$ and ${}^3He-B$ in $d=3$,  in which the infrared {\em wavefunctions } in the bulk are known to be identical  to the Euclidean  {\em correlation functions } at the boundary. We show  how this comes about in detail. By extending our logic (which requires only Lorentz invariance)   to $d=3$ we explain  the same equality  for ${}^3He-B$. While our tactic is most readily demonstrated by these two examples, it is not restricted to them.
\end{abstract}

\maketitle

The boundaries or edges of condensed matter systems received  scant attention until  recent developments showed them to be fertile areas of research both in the  Fractional Quantum Hall Effect (FQHE) \cite{halperin,wen}.
and  in topological  insulators and superconductors \cite{haldane0,kane,jmoore,zhang,roy,ludwigTI,kitaev0}.

In two spatial dimensions, the edge dynamics is described  by conformal  field theory \cite{wen} which was also used
  to produce  wave functions in the bulk \cite{greiter,mooreread}.   Moore and Read\cite{mooreread} showed that one may view the FQHE  wavefunctions and the quasi-hole excitations as  conformal blocks in which  both electrons and the quasiparticle coordinates  are treated on the same footing and their charges and braiding properties are severely constrained.  For an exhaustive review of many related topics  see Nayak {\em et al} \cite{nayak}.

What are the minimal ingredients necessary to establish equality of edge
correlations and bulk wavefunctions? Are analytic functions or d=2 conformal invariance required? We show that our edge-bulk  equality follows for a class of topological superconductors in various dimensions invoking {\em only} approximate
Lorentz symmetry. The
connections obtained here using an effective low energy hamiltonian differ from CS
theory\cite{witten} in which the hamiltonian vanishes and only non-dynamical particles enter via Wilson loops, as reviewed in Ref.\onlinecite{nayak}.

 %Here we {\em explain} the observed {\em equality} (in the infrared) of the $N$-body bulk wavefunctions and $N$-point edge correlation function for two illustrative examples, the   $p+ip$ superconductor in $d=2$ and superfluid ${}^3He-B$ in $d=3$ using  only  approximate Lorentz invariance of the  action and also apply  it to  fractionalized systems.
%The connections obtained here using the appropriate low energy hamiltonian differ  from CS theory \cite{witten}  in which the  hamiltonian vanishes and  only  non-dynamical particles  enter via Wilson loops,  as reviewed in Refs. \onlinecite{nayak}.

We shall first write down an operator expression for  $Z(J)$, the generating function of $N$-body wavefunctions of the  bulk. This is shown to be accomplished by introducing a time dependent chemical potential that changes abruptly at some Euclidean time. We then drop some high derivative terms which do not matter in the infrared, and express $Z(J)$ as a Grassmann integral over a Lorentz invariant action.
Rotating by 90 degrees to exchange time and a  spatial    direction
we obtain the {\em same} topological superconductor but with a spatial edge induced by the jump in chemical potential.
We find that the same $Z(J)$ has now morphed into  the generating function for the correlation functions of the edge excitations. Three examples are given: the  $p+ip$ superconductor in $D=2+1$,  ${}^3$ He B phase in $D=3+1
$  and a p-wave superconductor (the Ising model) in $D=1+1$.

{\em Extracting Wavefunctions:}  Recall that given    a second-quantized  $N$-body state $|\Phi \rangle$ with wavefunction $\phi (x_1, x_2, .. x_N)$
we  extract $\phi$ using
 \beq
  \phi (x_1, x_2, .. x_N) = \langle \emptyset |\Psi (x_1) ... \Psi (x_N)|\Phi \rangle . \label{wavefunction}
 \eeq
 where $\langle \emptyset|$ is the Fock vacuum and $\Psi$ is the canonical electron destruction operator.
For problems with variable number of particles, let us  define  the
generating function
 \beqr
  Z(J)
  &=& \langle \emptyset |e^{\int dxJ(x)\Psi (x) }|\Phi \rangle \label{zj1}
  \eeqr
  which yields  $N$-body wavefunctions upon differentiating  $N$- times with respect to the Grassmann source $J(x)$.

%We have in mind the specific case
%\beq
%|\Phi\rangle =|BCS\rangle= \exp \left(\half  \int %\Psi^{\dag}(x)g(x-y)\Psi^{\dag}(y)dx dy\right)\label{bcsgs}
%\eeq
%where $g$ is the pair wavefunction and $x$ and $y$ stand for all spatial coordinates,
We want to express $Z(J)$ as a path integral when $|\Phi\rangle$ is the ground state of a Hamiltonian $H$ without conserved particle number. Since  Euclidean time evolution for long times projects to the ground state,   we can obtain  $|\Phi \rangle$ as
   %from  $\tau=-\infty $ to $x_3=0^-$
  \beq
  |\Phi \rangle = U(0^-, -\infty )|i\rangle
  \eeq
  where $|i\rangle$ is a generic initial state and $U(0^-, -\infty )$ is the imaginary time propagator from $-\infty$ to $0^-$. Then we insert the operator $\exp \left[ \int J(x) \Psi (x)dx\right] $ at time $0$.
  Finally, we obtain the Fock vacuum by evolving a generic state $\langle f|$ from time $+\infty$ to $0^+$ using a hamiltonian $H'$ with a huge negative $\mu$ that empties out fermions so that we may write  $\langle \emptyset |=\langle f|U(\infty,0^+)$. Thus
\begin{equation}
Z(J)= \langle f|U(\infty,0^+) e^{\int J(x) \Psi (x)dx} U(0^-, -\infty )|i\rangle
\end{equation}
which has a path integral representation.

{\em Example 1: $p+ip $ :}  The mean-field  hamiltonian  is\cite{volovik,readgreen}:
 \beq
  H=\sum_k (c^{\dag}_{k}, c_{-k}) \left(
                           \begin{array}{cc}
                             \al k^2 - \mu & \Delta \cdot (k_1-ik_2) \\
                             \Delta^*\cdot (k_1+ik_2)  & -(\al k^2 - \mu) \\
                           \end{array}
                         \right)\left(
                                  \begin{array}{c}
                                    c_k \\
                                    c^{\dag}_{-k} \\
                                  \end{array}
                                \right)\label{bcsh}
  \eeq
  here $1,\, 2$ are spatial indices and $x_3$ will be time. We employ the minimum  $k$ dependence in the pairing function, and set the coefficient $\Delta=1$ for convenience so the gap function is:
  %and set them equal to unity so that  our gap function is simply
%  \beq
$  \Delta (k_1,k_2) = k_1- ik_2$
%\eeq

The $\alpha  k^2$ term is needed to ensure the nontrivial  topology of the weak-coupling phase\cite{readgreen} and to populate it with electrons for $\mu >0$. We shall remember this association but drop the `$k^2$' term in the computations since it does not affect infrared correlations.

  Now the  mean field Hamiltonian in real space:
  \beq
  H= \!\!\! \int \! d^2x \left[ \Psi^{\dag} ( -\mu)  \Psi + \half
  (\Psi^{\dag } (-i\partial_1 - \partial_2)\Psi^{\dag}+ h.c)\right].\label{bdg}
  \eeq
leads to  corresponding
     Grassmann action  for $U(0,-\infty )$: \beqr
   S&=& \!\!\!\int_{-\infty}^{\infty} d^2x\int_{-\infty}^{0}dx_3\!  \left[\bar{\psi } {\cal D} \psi +\bar{\psi}i\partial \bar{\psi}+  \psi i\bar{\partial}\psi \right]\ \ \ \label{grassact2}\\
   {\cal D}&=& (-\partial_3 + \mu )\ \ \ \ \ \
   \partial = {\partial\over \partial z}\ \ \ \ \   \bar{\partial} = {\partial\over \partial\bar{ z}}\label{grassact3}
   \eeqr
For the  $0^+<x_3<\infty$, we choose  $\mu = \mu_{+}$,  a very large negative number, associated with the Fock  vacuum and obtain, for all $x_3$, the   action including the source $J$:

   \beqr
   S(J)&=&  \!\!\! \!\! \int_{-\infty}^{\infty}\!\!\! d^3x\!\! \left[\bar{\psi } {\cal D}\psi + \bar{\psi}i\partial \bar{\psi}+  \psi i\bar{\partial}\psi + J\psi \delta(x_3) \right] \label{grassact4}
   \eeqr
   where ${\cal D}$ now contains a time-dependent $\mu (x_3)$ that jumps at $x_3=0$ from $\mu_->0 $ to $\mu_+ \rightarrow -\infty $.

   The generating function of the BCS wavefunctions  is
   \beq
   Z(J)= {\int \left[  d\bar{\psi}d\psi\right]e^{S(J)}\over \int \left[  d\bar{\psi}d\psi\right]e^{S(0)}}\label{generating}
   \eeq

\begin{figure}
\includegraphics[width=8cm]{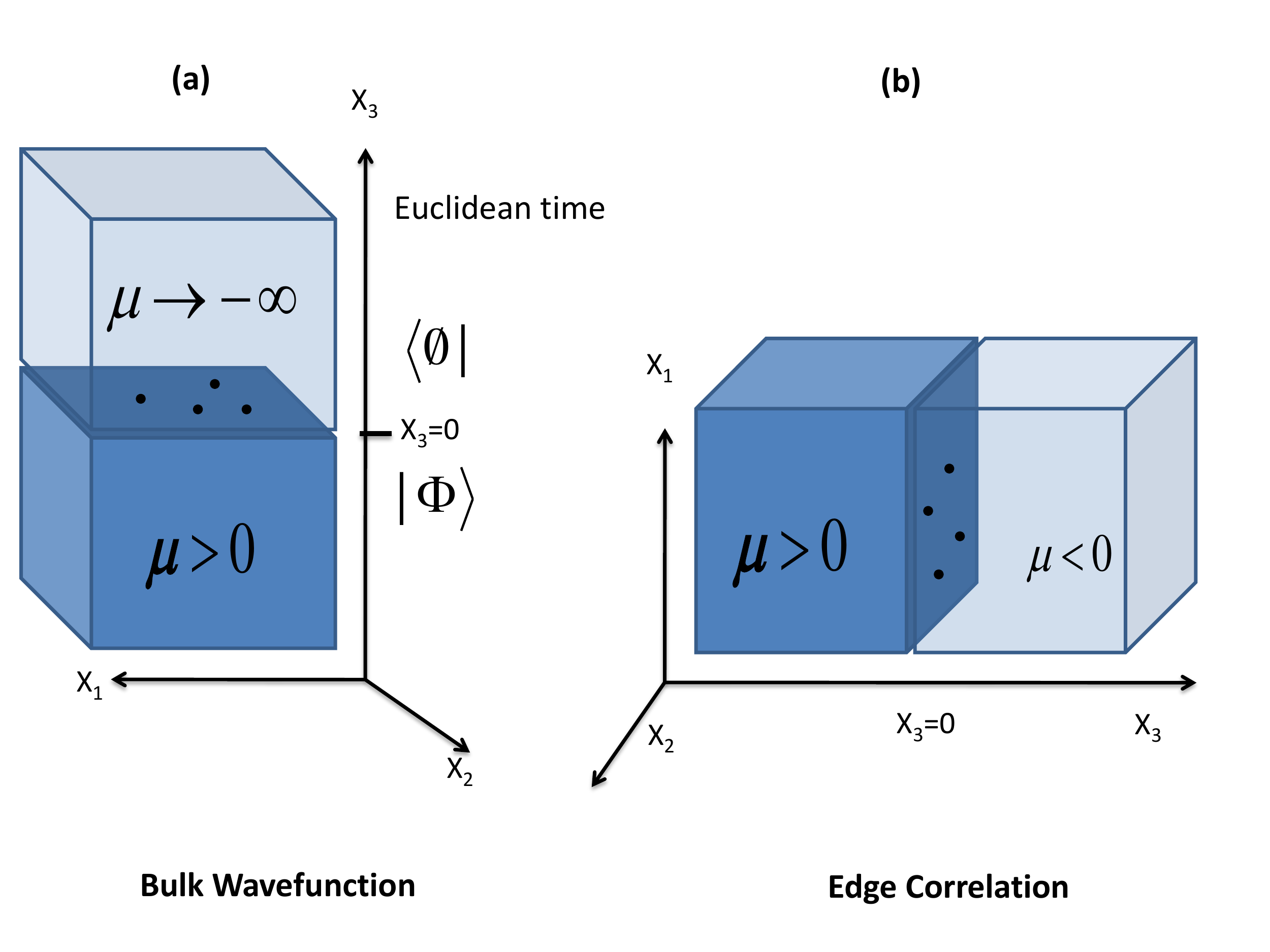}
\caption{(a) Wavefunction: The original superconductor with $\mu = \mu_->0$ lies in the $x_1-x_2$ plane and evolves in Euclidean time $x_3$ from $-\infty$ to $0^-$, projecting out the ground state $|\Phi\rangle$.  At $x_3=0^+$ the  chemical potential drops abruptly to a large negative value $\mu^-$, leading to the Fock vacuum. (b) Correlation functions: A Lorentz rotation makes  $x_1$  the new time and $x_3$ a the spatial coordinate along  which the system has an edge at $x_3=0$. The world-sheet  of the edge lies in the $x_1-x_2$ plane at $x_3=0$.\label{edje}}
\end{figure}

The story is depicted in the left half of Figure \ref{edje}:
the fermions   travel unsuspectingly along in Euclidean time $x_3$ and slam like bugs onto the windshield at $x_3=0^{-}$ when  $\delta(x_3) J \psi$  kills      them.

Since  $\psi$ and $\bar{\psi}$ in Eq. \ref{grassact4} are independent Grassmann variables,  we integrate out $\bar{\psi}$  to obtain the effective action for just $\psi$ to which alone $J$ couples:
   \beqr
   S_{eff}(\psi, J)&=& \int d^3x \left(\psi i\bar{\partial}\psi +J\psi  +  \psi {1 \over 4i\partial}{\cal D}^T{\cal D} \psi\right)  \nonumber\\
   &\equiv &S_0(J) + S_{ind}. \label{effecS}
   \eeqr

For the infrared limit we keep just the Jackiw-Rebbi zero mode \cite{jackiw} of the hermitian operator
   \beq
  {\cal D}^T{\cal D}(x_3)=  (\partial_3 + \mu (x_3))(-\partial_3 + \mu (x_3)),\label{h3}
   \eeq
 that obeys  ${\cal D}f_0=0$
   \beq
   f_0 (x_3) = f_0(0)  e^{\int^{x_3}_{0} \mu (x')dx'}
   \eeq
 in the mode expansion of the Grassmann field:
   \beq
   \psi(x_1,x_2,x_3) = f_{0}(x_3)\psi(x_1,x_2).
   \eeq
   This kills $S_{ind}$, and upon integrating $f_{0}^{2}$ over $x_3$,
   \beqr
   S_{eff}(J) &=& \!\!\int dx_1dx_2 \,\psi  (i\bar{\partial}  +  J f_0(0) )\psi \label{seff}
   \eeqr
 While this is indeed  the action of a chiral majorana fermion living in the $1-2$ plane
we are not done:  we need to show that this  fermion and this action also arise at the edge of the  {\em same } $p+ip$ system. But so far we have no edge! It will be introduced shortly, but first a summary of results on the wavefunction.

{\em Pfaffian Wavefunction:}  Integrating over $\psi$ in Eq. \ref{seff},  and  suppressing  the constant $ f_{0}^{2}(0)) $ we find
  \beqr
  Z(J) &=& \exp \left[{\int d^2  \br  J (\br ) \left[{1 \over 4i \bar{\partial}}\right]_{\br \br'}J (\br') }\right]\eeqr

 The two-particle wavefunction $\phi (\br_1 -\br_2)$ can be  written in terms of many related quantities:
\beqr
\phi = {\partial^2  Z(J)\over \partial J_1 \partial J_2} \!\! &=& \!\!\! \left[{1 \over 2i \bar{\partial}}\right]_{\br_1 \br_2}\!\!\!\!=\!  \Delta^{*-1}_{\br_1 \br_2}={1 \over z_1-z_2}
\eeqr
and the $N$-particle wavefunction is  ${\rm Pf}({1 \over z_i-z_j})$. In the Supplementary Material we relate $Z(J)$ and the conventional BCS wavefunction:
\beq
|BCS\rangle= \exp \left(\half  \int \Psi^{\dag}(x)g(x-y)\Psi^{\dag}(y)dx dy\right)\label{bcsgs}|\emptyset\rangle
\eeq
and see that  $\phi=-g(\br_1-\br_2)$.

{\em The Edge:} To relate $Z(J)$  in Eqn. \ref{grassact4}  to a problem with the  edge  we rewrite $S(J)$  in Lorentz invariant form:

\beqr
S(J)
&=&\int d^3 x \left[ \bar{\Psi}\left(\partial \!\!\!/ -\mu\right)\Psi + J^T \Psi\right]\ \ \mbox{where }\label{linva2}\\
\Psi &=& \left(
         \begin{array}{c}
           \psi \\
           \bar{\psi} \\
         \end{array}
       \right)\ \ 
\bar{\Psi}= \Psi^T \e; \ \ \e=i\s_2 \ \ \ \partial \!\!\!/= \gamma_{\mu}\partial_{\mu} \label{linva3}\\
\label{linva4}
\gamma_1 &=& \s_2 \ \ \ \gamma_2= -\s_1 \ \ \ \ \gamma_3 = \s_3\label{lia3}\\
J^T&=&J \delta(x_3)(1\,0).\label{linva5}
\eeqr

Look at the left half of Figure 1.  We see our current description of the superconductor:   translationally invariant in  the $x_1-x_2$ plane, regarded as the space in which the $p_1+ip_2$ superconductor lives,  and with  a jump in $\mu$ at "time"  $x_3=0$.    In this description, the functional integral is saturated by  one mode $f_0(x_3) $, glued to the interface, exactly like the electron gas at a heterojunction.

Extracting $H_{}(x_1,x_2)$ from the Lorentz invariant action is like taking the row-to row transfer matrix.
To
derive  the hamiltonian that governs the column-to -column dynamics,  we rotate  the three dimensional spacetime by $-{\pi \over 2}$ around the $x_2$ axis to obtain the view shown in the right half of  Figure \ref{edje}. The points carry the same labels as before but the  spinor undergoes a rotation:
\beq
\Psi= \left(
  \begin{array}{c}
    \psi \\
    \bar{\psi} \\
  \end{array}
\right)
= e^{ i{\pi \over 4}i\gamma_3\gamma_1}\left(
                                \begin{array}{c}
                                  \psi' \\
                                  \bar{\psi}' \\
                                \end{array}
                              \right)
= e^{ i{\pi \over 4}\s_1}\Psi'\label{rot}
\eeq
Upon performing this transformation we end up with
\beqr
S(\Psi',J)&=& \!\!\!\!\int d^3x\left[\bar{\Psi}'\left[ \s_3\p_1 \!-\s_1\p_2\!-\s_2\p_3\!-\mu\right]\Psi'\!\nonumber \right.\\&+&\left. J\delta(x_3) ({\psi'+i \bar{\psi}'\over \sqrt{2}})\right]\label{3d1}
\eeqr
which describes exactly the same $p+ip$ superconductor but in the $2-3$ plane (with $1 \to 3, 3\to -1$) with an edge at $x_3=0$. An  $\al (k_{2}^{2}+ k_{3}^{2})$ term may now be added without affecting infrared edge correlations. This is required to complete our identification of regions with (and without)  fermions with $\mu$ positive (negative).

To see that the field   ${\psi'+i \bar{\psi}'\over \sqrt{2}}$ that $J$ couples to is precisely the  Majorana field that arises at the edge, consider  solving the equation for the zero mode which follows  from Eq. \ref{3d1}  on dropping all $x_1$, $x_2$ dependence:

\beq
(\s_2\p_3+ \mu (x_3))\chi_0' =0 \,\Rightarrow \,\chi_{0}' (x_3)={ 1\over \sqrt{2}}\left(\begin{array}{c}
                                   1 \\
                                   -i
                                 \end{array}\right) f_{0}(x_3).\label{edgemaj2}
       \eeq
%       with  
%       \beq
%       \chi_{0}' (x_3)={ 1\over \sqrt{2}}\left(\begin{array}{c}
%                                   1 \\
%                                   -i
%                                 \end{array}\right) f_{0}(x_3).\label{edgemaj2}
%                                \eeq
the normalizable spinor solution indeed corresponds to   the operator ${1 \over \sqrt{2}}(\psi' + i \psi'^{\dag})$.

We are done, for we have shown that $Z(J)$ is at once the generators of electronic wavefunction in the bulk and of correlation functions of the Majorana field at the edge.

For completeness,  the edge Majorana field action follows from saturating the   $x_3$ dependence of $\Psi'$ as follows:
\beq
\Psi'(x_1,x_2,x_3) = { 1\over \sqrt{2}}\left(\begin{array}{c}
                                   1 \\
                                   -i
                                 \end{array}\right) f_{0}(x_3)
                                  \psi'(x_1,x_2)
                                  \eeq
                                  Plugging this into the action $S(\Psi',J)$  one finds, upon integrating the normalized function $f_{0}^{'}(x_3)$ over $x_3$
                                  \beq
                                  S(\Psi',J)\to \int dx_1 dx_2 \left[\psi'i\bar{\partial}\psi'+J f_{0}(0)\psi'\right]
                                  \eeq
exactly as in Eqn.\ref{seff}, for the wavefunction.

{\em Example 2: ${}^3He-B $ in D=3+1:} In a simplified model of superfluid ${}^3He-B$, Cooper pairs have spin $1$, whose projection lies perpendicular to the momenta $\pm{\bf k}$ \cite{leg,vw}. The winding of this axis around the Fermi surface in the weak pairing phase leads to its topological properties\cite{vw,jdp}. The mean-field Hamiltonian for this time-reversal invariant class DIII system is \cite{leg,vw} is:
%in natural units is
\beqr
\label{HMF}
 H &=& \sum_{{\bf p}\sigma\sigma'}\Psi^\dagger_{{\bf p}\sigma}(\frac{{\bf k}^2}{2m}-\mu)\Psi_{{\bf k}\sigma}\\\nonumber
 &&+\left \{\Delta_{{\bf k}\sigma\sigma'}\psi_{{\bf k}\sigma}\psi_{-{\bf k}\sigma'} + {\rm h.c.}\right \}\\\nonumber
 \Delta_{{\bf k}\sigma\sigma'} &=& [\e \bk \cdot \sigmab ]_{\sigma\sigma'}
\eeqr

The   $d=3$ problem is just the $d=2$ problem on steroids: $\Delta$  goes from being a complex number to a quaternion,   and the spinless fermion    is replaced by   a two-component spinor.
Hence the weak-pairing wavefunction
 is $g_{\sigma_i\sigma_j}(\br_{ij}) \sim \frac{[ \br_{ij}\cdot \sigmab \e]_{\sigma_i\sigma_j}}{r^3_{ij}}$ and the many-body wavefunction is the corresponding ${\rm Pf} (g)$ as noted in Ref. \onlinecite{ludwigTI}  .

 The Lorentz invariant action for the wavefunction  is
\beqr
S&=& \int d^4x \half \bar{\Psi}\left[ \partial\!\!\!/ - \mu\right]\Psi\ \ \ \ \mbox{where}\\
\gamma_0 &=& \left(
             \begin{array}{cc}
               I & 0 \\
               0& -I \\
             \end{array}
           \right)\ \ \ \ \   \gammab = \left(
                                         \begin{array}{cc}
                                           0 & i  \sigmab \e\\
                                           i\e \sigmab & 0 \\
                                         \end{array}
                                       \right)\\
                                       \bar{\Psi}&=&  \Psi^T\left(
                                                                                   \begin{array}{cc}
                                                                                     0 & I \\
                                                                                     -I& 0 \\
                                                                                   \end{array}
                                                                                 \right)
                                        \label{lia4d}\eeqr

Now the $0$ and $1$ directions are exchanged by
                                        $
                                        R= \exp {\left[{i\pi \over 2}{i\gamma_0\gamma_1 \over 2}\right]}$, so that
 $J$ now couples to ${\psi' +i\s_3\psi'^{\dag}\over \sqrt{2}}
                                                      $
                                                    which is readily verified, as before,  to be  the gapless edge mode of the rotated  theory.
                                                          The action for the edge theory obtained by saturating with the zero mode is
                                                          \beq
                                                          S_{edge}= \int d^3x \half \bar{\psi} \partial \!\!\! / \psi\ \ \ \ \
\partial \!\!\! /=\sigma_{j}^{} \partial_j\ \ \ \
\bar{\psi}= \psi^T (-\sigma_2)
\eeq

{\em Example 3:} We could equally well go {\em down} a dimension, to a spinless p-wave superconductor in $d=1+1$\cite{kitaev}  where  $\Delta = k_x$, which is also related to the quantum Ising model, via the Jordan-Wigner mapping.  The edge theory is  $0+1$ dimensional, corresponding to a Majorana zero mode, with  Lagrangian ${\cal L} = \half \psi \partial_x \psi$.

{ \em Fractionalized Topological Superconductors:}
 We construct a fractionalized superconducting phase in D=2+1 that  bears the same relation to the $p+ip$ superconductor as the Laughlin $m=3$ quantum Hall state bears to the integer Hall effect. Consider splitting the electron operator at each site into three fermions ('partons')  $c_r=if_{1r}f_{2r}f_{3r}$ and $c^\dagger_r=if^\dagger_{1r}f^\dagger_{2r}f^\dagger_{3r}$ \cite{Wen, Zhang} with
  the following $p+ip$ mean field action for the partons:
 \beqr
 \label{partons}
   S(J) &=& \int d^3x \left [ {\mathcal L}_0 + iJf_{1}f_2f_3\right ]\\
\nonumber
   {\mathcal L}_0 &=& \half \sum_{a=1}^3\left(  \bar{f}_a\  f_a \right)\left(
                                   \begin{array}{cc}
                                     -\partial_3 +\mu & i\partial_1+\partial_2    \\
                                      i\partial_1-\partial_2 &  -\partial_3 -\mu \\
                                   \end{array}
                                 \right)\left(
                                          \begin{array}{c}
                                            f_a \\
                                            \bar{f}_a \\
                                          \end{array}
                                        \right)
                                        \eeqr
 When the gauge theory is in a deconfined phase, the partons accurately describe the low energy dynamics. The SO(3) symmetry of the action, a remnant of the SU(3) gauge redundancy  implied by $c_r=if_{1r}f_{2r}f_{3r}$ \cite{Wen1}, is the gauge symmetry here. When the gapped  bulk is integrated out, it generates  an SO(3)$_1$ (or equivalently SU(2)$_2$)\cite{Nayak} Chern-Simons term which renders the gauge field massive thereby liberating the partons with the action in Eq. \ref{partons}.

  Emptying out the  electrons requires removing the $f$ fermions, hence the strong pairing phase of the $f$s, where their chemical potential is taken to be large and negative, corresponds to the Fock vacuum.  The  electron correlators involve products of three parton correlators each  in a p+ip state, so the electronic wavefunction is: 
\beq
\Psi(z_1,z_2,\dots, z_{2N})=\left \{ {\rm Pf}\left [ \frac1{z_i-z_j}\right ]\right \}^3
\eeq

Equivalently we can start from the edge where the three Majorana modes are massless by gauge symmetry and have no relevant short range interactions in three spacetime dimensions. Long range gauge interactions do not exist  due to the Chern-Simons term. Consequently the bulk effective action must also be described by three noninteracting fermions.

{\em Summary:} We have explained why  the electronic wavefunctions in  the bulk coincided with the massless Majorana  correlation functions at the edge in certain problems.
We first wrote  $
    Z(J)= \langle \emptyset |e^{J\Psi} |BCS\rangle$   as a path integral in which the chemical potential abruptly jumped at in Euclidean  time. Dropping the `$k^2$' terms which determined boundary conditions on $\mu$, we obtained a Lorentz invariant action.
     Upon rotation by $\pi/2$ the same action  described a system that had an edge and  $Z(J)$ had meanwhile   morphed into the generating function for edge correlations.
In general, rotating axes will relate bulk wavefunctions to the edge correlations of a  different (possibly unnatural) problem. The examples considered here are self-dual in this respect.

     Our analysis holds in many dimensions and applies to fractionalized cases as well, as long as   varying $\mu$ can  change the topology. This is possible in the Altland-Zirnbauer classification\cite{ludwigTI} for  models in class D  in d=1 and d=2 (like p+ip), in class C in d=2 (like d+id)  and class DIII in d=2,3 (He-3 B phase) but not for  classes like CI in d=3,\cite{RyuLudwig}  which additionally rely on band topology of the weak pairing Fermi surface.
      We are currently modifying  our  derivation for Laughlin quantum Hall states, where   $\mu$  couples to a conserved charge.

 The entanglement spectrum of the bulk seems to determine the  edge theory  \cite{haldane,others},  which we now   relate back to the bulk wavefunction. Since the entanglement of a gapped phase appears from near the cut, {\em  the entire bulk wavefunction must be coded  holographically  in  every   $d-1$ dimensional sliver probed in the entanglement analysis.}  
 
 Previously, the connection between edge
states and bulk wavefunctions has played an important role identifying new FQH
states\cite{mooreread, readrezayi}. Our work suggests a similar approach could be
fruitful in identifying interacting
topological phases in D=3+1.

We thank  the  NSF for grants DMR- 0645691 (AV), and DMR-
0103639 (RS) and   Tarun Grover, Greg Moore   and Shoucheng Zhang for detailed suggestions. RS thanks the Department of Physics at UC Berkeley  for  hosting his sabbatical  in Spring 2011.

\newpage
\section{Supplementary Material}
Suppose  we are given a Majorana hamiltonian
\beq
H= \half  \sum_{ij}\Psi_i h_{ij}\Psi_j
\eeq
wherein \beq
\{ \Psi_j,\Psi_j\}= \delta_{ij},
\eeq
where  $i,j$ subsume all labels, spatial and internal. If the labels are continuous, the Dirac $\delta$ should be used   and derivatives $\partial$  viewed as  antisymmetric matrices.

By definition the Grassmann integrals are
\beq
\int \psi d\psi =1 \ \ \ \  \int 1 \cdot d\ \psi =0
\eeq

The Euclidean  path integral corresponding to $h$ is
\beq
Z= \int [  d{\psi}] e^{ \half \int  dt \sum_{ij}\psi_i  (t) (-\partial_t \delta_{ij} - h_{ij})\psi_j (t)}
\eeq

Since coherent states do not exist for Majorana operators (which square to $\half$ and not $0$), one way to derive this result is to first form Dirac operators from pairs of Majorana operators, use fermion coherent states for the former to obtain a path integral and then undo the transformation back to Majorana fields.

The Gaussian integral that is repeatedly used is
\beq
Z(J)= \int  e^{\half \chi A \chi+J \chi}[d\chi]=Pf (A) e^{\half J A^{-1}J}
\eeq
where $J$ and $\chi$ are $2N$ -component Grassmann vectors.

The two-point correlator is
\beq
\langle \psi_a \psi_b \rangle = \left. {\partial^2 Z(J)\over \partial J_a \partial J_b}\right|_{J=0}= A^{-1}_{ba}=-A^{-1}_{ab}
\eeq
The minus sign in the last term can be avoided if  the exponent is written as $e^{-\half \chi A \chi}$.

Higher correlators are given by Pfaffians.
\subsection{Pfaffian wavefunctions}
Let us put these ideas to work in deriving the many-body wave functions from the second quantized BCS state.

Consider the generating function of  wavefunctions for any number of particles
  from which the wavefunctions can be obtained by differentiating with respect to the Grassmann source $J(x)$
 \beqr
 \lefteqn{Z(J) =}\nonumber \\& & \langle \emptyset |e^{\int dx J(x)\Psi(x)}|BCS\rangle \nonumber \\
 &=& \langle \emptyset | e^{\int dx J(x)\Psi(x)}\cdot I  \cdot e^{ \half \int \Psi^{\dag}(x) g  (x-y) \Psi^{\dag}(y) dx dy} |\emptyset \rangle \nonumber \\
 &\equiv & \int \left[ d\bar{\psi} d \psi\right] e^{-\bar{\psi} \psi }\langle \emptyset |e^{ J(x)\Psi} |\psi\rangle \langle \bar{\psi} |e^{ \half \Psi^{\dag} g   \Psi^{\dag}}|\emptyset \rangle \label{grass1}
 \eeqr
 where, in the last step we have resorted to a compact notation and inserted the following
    resolution of the identity in terms of Grassmann coherent states:
 \beq
 I = \int |\psi \rangle \langle \bar{\psi} |e^{-\bar{\psi}  \psi } \left[ d\bar{\psi} d \psi\right]
 \eeq
 and where it is understood for example that
 \beq
|\psi \rangle = \prod_{x} | \psi (x)\rangle \ \ \ \ \ \   \left[ d\bar{\psi} d \psi\right]= \prod_{x}\left[ d\bar{\psi} (x)  d \psi (x) \right]
\eeq
It is important to remember that $\bar{\psi}$ and $\psi$ are independent and dummy variables. Using the defining property of coherent states
\beq
\Psi | \psi \rangle = \psi |\psi \rangle \ \ \ \  \langle \bar{\psi} |\Psi^{\dag}= \langle \bar{\psi} |\bar{\psi} \eeq
 in Eq. \ref{grass1} we find
 \beq
  Z(J)=\int \left[ d\bar{\psi} d \psi\right] e^{-\bar{\psi} \psi }e^{J\psi}   e^{ \half \bar{\psi} g   \bar{\psi}}\label{grass2}
  \eeq
  where we have used the fact that
  \beq
  \langle \emptyset |\psi\rangle \langle \bar{\psi} |\emptyset \rangle =1
  \eeq
  since at each site
  \beq
  |\psi \rangle  = |0\rangle - \psi |1\rangle \ \ \ \  \langle \bar{\psi} | = \langle 0| - \langle 1|\bar{\psi}
  \eeq
  and $|\emptyset \rangle = |0\rangle \otimes |0\rangle \otimes ....|0\rangle$.
Doing the integrals over $\psi$ and $\bar{\psi}$, we find
  \beq
  Z(J) = e^{\half J g J}\label{zj}.
  \eeq
  The pair wavefunction is
 \beq
  \phi(x_1,x_2)=\left. {\partial^2 Z\over \p J(x_1) \p J(x_2) }\right|_{J=0}= -g(x_1-x_2)\label{2pc}
  \eeq

  Higher correlations follow from Wick's theorem. For example
  \beqr
  \phi (x_1,x_2,x_3,x_4)&=&  g(x_1-x_2) g(x_3-x_4)\nonumber \\&-& g(x_1-x_3) g(x_2-x_4)\nonumber \\
   &+&  g(x_1-x_4) g(x_2-x_3).\label{4pc}
  \eeqr

 \end{document}